Choosing an algorithmic fairness metric for an online marketplace:

Detecting and quantifying algorithmic bias on LinkedIn

YinYin Yu*

LinkedIn, yinyin.yu14@gmail.com

Guillaume Saint-Jacques

Apple, guillaume.saintjacques@gmail.com

In this paper, we derive an algorithmic fairness metric from the fairness notion of *equal opportunity for equally qualified candidates* for recommendation algorithms commonly used by two-sided marketplaces. We borrow from the economic literature on discrimination to arrive at a test for detecting bias that is solely attributable to the algorithm, as opposed to other sources such as societal inequality or human bias on the part of platform users. We use the proposed method to measure and quantify algorithmic bias with respect to gender of two algorithms used by LinkedIn—a popular online platform used by job seekers and employers. Moreover, we introduce a framework and the rationale for distinguishing algorithmic bias from human bias, both of which can potentially exist on a two-sided platform where algorithms make recommendations to human users. Finally, we discuss the shortcomings of a few other common algorithmic fairness metrics and why they do not capture the fairness notion of equal opportunity for equally qualified candidates.

## 1 INTRODUCTION

Algorithmic fairness is an increasingly important issue as AI-powered systems become more prevalent in our everyday lives. Some examples of substantial decisions influenced by algorithms include setting bail [4], granting loan applications [3], college admission [5], and hiring [1]. The algorithmic fairness literature has entertained a diverse set of algorithmic fairness metrics. While many papers abstracted from discussing why one metric was chosen above others, others have pointed out that different metrics hold conflicting definitions of fairness and attempted to assess the pros and cons of different definitions. In this paper, we take a different approach; we begin with a fairness notion, *equal opportunity for equally qualified candidates, regardless of their protected categories* [8], from which we derive an algorithmic fairness metric that detects violations of this fairness notion[1]. We then apply this metric to detect and quantify algorithmic bias with respect to gender of two recommendation algorithms used on LinkedIn. We introduce a framework and discuss the rationale for separating the measurement of bias on the part of the recommendation algorithm from other sources of bias, such as human bias, on two-sided platforms where both the algorithm and human actors play a role in the arriving at the final outcome. Finally, we discuss other popular algorithmic fairness metrics and why they are incongruent with the fairness notion of equal opportunity for equally qualified candidates.

---

* All work was completed while the authors were employed by LinkedIn.

[1] Throughout this paper, we use the term "fairness" to mean adherence to the fairness notion of equal opportunity for equally qualified candidates, and, conversely, we use the term "bias" to mean violations of the said fairness notion.

## 2 BACKGROUND

### 2.1 About LinkedIn

In this paper, we propose a way to measure and quantify algorithmic bias based on the fairness notion of *equal opportunity for equally qualified candidates*. We showcase our methodology by applying our metric to data from LinkedIn—an online platform designed for professional networking, job seeking, recruiting, and career development, with over 830 million members across 200 countries and territories. LinkedIn enables registered members to, among other things, connect with other professionals, distribute and consume content, and search for jobs. It also hosts a myriad of functionalities created to meet the needs of employers, including job posting, candidate search, and event hosting, among other features.

Like most online platforms, LinkedIn is powered by a set of algorithms that helps users parse through the vast quantity of information that exists on the platform. For example, the People You May Know (PYMK) algorithm helps members grow their network by recommending potential professional connections. The InstaJob algorithm notifies job seekers of certain newly posted jobs to which they may be interested in applying. The Recruiter Search algorithm returns a ranked list of relevant candidates for a keyword search initiated by a recruiter. The Feed algorithm helps members sort relevant content from their professional networks. These are just a few examples of the many algorithms that power LinkedIn.

### 2.2 Recommendation algorithms

Recommendation algorithms, a common type of algorithms that often power two-sided platforms, will be the focus of this paper. A recommendation algorithm makes recommendations to human decision makers who have the ultimate agency in deciding whether to act on the recommended entities. For example, LinkedIn's InstaJob algorithm recommends certain newly posted jobs to job seekers and encourages them to apply, but the decision of whether to apply ultimately lies in the hands of the job seekers[2]. LinkedIn's PYMK algorithm recommends potential connections to help broaden members' networks, but members decide whether to act on the recommendations and connect with each other. We refer to the party for whom the recommendations are made as the "viewer" and the recommended entities as the "candidates". The success of a recommendation algorithm is evidenced by the actions that the viewer takes with respect to the candidates, i.e., a recommendation is successful if it is adopted by the viewer. Therefore, a recommendation algorithm's objective is usually a function of actions taken with respect to the recommendations. For example, the InstaJob algorithm scores job-member pairs based on the predicted value of an objective that is a function the member's application decision and the attention his/her application receives from recruiters if the member were notified about the job:

$$Y_i = \alpha * 1(applied\ to\ job\ \&\ received\ no\ recruiter\ attention\ |\ notified)_i$$
$$+ 1(applied\ to\ job\ \&\ received\ recruiter\ attention|\ notified)_i$$

---

[2] Note that, aside from receiving an InstaJob notification, members can also discover these jobs on or off LinkedIn's platform, including through LinkedIn's search functionality.

where $i$ is the job-candidate subscript, and $\alpha \in (0,1)$ is a (proprietary) scalar parameter that determines how the algorithm rewards applications with recruiter attention relative to those without[3]. Notifications are sent to candidates whose predicted objective value is greater than a common threshold. Similarly, the PYMK algorithm scores and ranks candidate members based on the predicted value of their propensity to connect with the viewer member and engage in downstream interactions conditional on connecting. The exact form of the proprietary objective is redacted.

### 2.3 Notations and data generating process

We will focus on algorithmic fairness of pointwise recommendation algorithms with respect to the candidates' protected group membership in this paper; see [7] for an extension of the methodology discussed in this paper to listwise ranking algorithms[4]. A pointwise recommendation algorithm scores candidates based on predictions of its objective, $s_i \equiv s(X_i) \equiv E[Y_i|X_i]$[5], where $X_i$ is a set of features used to make prediction for candidate $i$. This score is in turn used to allocate opportunities. For classification algorithms, the treatment is allocated to those whose scores fall above a common threshold, $T_i = 1(s_i \geq \bar{S})$, and denied to those whose scores fall below the threshold score, $\bar{S}$. For ranking algorithms, candidates are ranked in decreasing order of their scores in a query, where a query is a call to the algorithm to generate a single ranked list of candidates for a viewer at a point in time[6]. In the context of ranking, we use $T_i \in \{0,1\}$ to denote whether the candidate has been impressed by the viewer[7]. Viewers scroll through the ranked list of candidates from top to bottom and can exit the list at any time. Therefore, an impression is not guaranteed for a particular candidate as the viewer may not scroll down far enough to view that candidate, and different viewers have different scroll depths. We will refer to the actual value of $Y_i$ that is realized for candidate $i$ after being impressed as $i$'s *realized outcome*; these are the actions that were taken with respect to the algorithm's recommendation $i$ conditional on impression.

We posit a data generating process where candidate's realized outcome is driven, probabilistically, by his/her unobserved true qualification, where qualification is defined with respect to the objective of the algorithm[8]. For example, in the context of InstaJobs, a more compatible job-candidate pair is more likely to result in the candidate responding to the InstaJobs notification with an application and the recruiter giving attention to the candidate's application. However, the scenario where a less compatible pair resulted in mutual interactions whereas a more compatible one did not is still possible due to the idiosyncrasies in the application and candidate evaluation process[9]. A corollary of this data generating process is that we cannot deduce the ranked order of candidates' unobserved true qualification by a single realization of their outcomes because being better qualified does not guarantee better outcomes. Formally, we model a candidate's realized outcome, $Y_i$, as a draw from a distribution parameterized by the candidates' true qualification, $F(.; q_i)$, where $F(.; q_1)$ first-order stochastically dominates $F(.; q_2)$ if $q_1 > q_2$. The condition of first-order stochastic dominance stipulates that better qualified candidates (higher $q_i$) are more likely to realize better outcomes (higher $Y_i$), although it is possible for a better qualified candidate to realize a worse outcome than a one who is less qualified in a single draw. Because different algorithms have different objectives/predict for different outcomes and qualification is defined with respect to an objective, we think of the $F(.; q_i)$'s as algorithm-specific, although the exact functional form of $F(.; q_i)$ is not important. Finally, we restrict our analysis to group-based fairness where candidates being classified/ranked belong to a set of mutually exclusive and collectively exhaustive groups, $G$, and group membership is observed.

---

[3] This is the objective of the version of the InstaJob algorithm that was audited. Newer versions of InstaJob algorithms may have a different objective.

[4] In a pointwise algorithm, the score is comparable across different queries, i.e., $s_i = s$ has the same interpretation, in terms of the algorithm's prediction of $Y$, across different queries. In contrast, the score is not comparable across queries in a listwise ranking algorithm; although each ranking is still created in decreasing order of the score, we can not interpret two candidates from different queries with the same score to mean that the algorithm has the same prediction for these candidates' $Y's$.

[5] The score of some pointwise algorithms may not on the same scale as $Y$, though the methodology discussed in this paper still holds. For ease of notation, we will assume $s(X) \equiv E[Y|X]$ without loss of generality.

[6] A query is a call to the algorithm to generate a single ranked list of candidates for a particular viewer at a point in time. For example, when a viewer navigates to the PYMK section of LinkedIn, the PYMK algorithm is called to generate a ranked list of candidates. If the viewer visits PYMK again at a later time, the algorithm is called to generate another ranked list.

[7] We say that a candidate has been impressed by the viewer if the candidate appears on the viewer's screen. Candidates for whom the viewer did not scroll down far enough to appear on his/her screen are said to be not impressed.

[8] Better qualified candidates are those more likely to realize higher values of the algorithm's objective, $Y_i$.

[9] The following thought experiment illustrates the probabilistic nature of the candidate evaluation process. Suppose N recruiters were to independently evaluate the same set of candidates for the same job. Different recruiters will most likely will arrive at slightly different stacked ranks of the candidates, but a more qualified candidate should, on average, be ranked higher than a less qualified one. Similar idiosyncrasies exist in the candidate application process.

## 3  TRANSLATING A FAIRNESS MISSION INTO A FAIRNESS METRIC

The algorithmic fairness literature has entertained a variety of conflicting fairness metrics, some of which cannot hold simultaneously except under special conditions [6]. Recent works have attempted to provide discussions on the pros of cons of different algorithmic fairness metrics [2], although in-depth discussions around why one fairness metric is used over another remains elusive in the algorithmic fairness literature. In this paper, we take a slightly different approach by deriving an algorithmic fairness metric from a fairness notion. We focus on recommendation algorithms commonly used on two-sided platforms such as LinkedIn and propose a couple of desirable properties that a fairness metric for this type of algorithms should possess; we also examine where other common fairness metrics are incompatible with our stated fairness notion.

### 3.1  Fairness notion

We derive a fairness metric based on the fairness notion of *equal opportunities for equally qualified candidates, regardless of their protected group membership* [8]. In addition, we propose a couple of desirable properties that an algorithmic fairness metric should possess,

    1. An algorithmic fairness metric should only capture the bias that the algorithm introduces onto the platform, on top of any other potential sources of bias; and

    2. An algorithmic fairness metric should be invariant to potential differences in preferences or qualification distributions between groups, much of which is unobserved and difficult to quantify.

The first criterion arises from the fact that, on a two-sided platform like LinkedIn, there are other actors at play, in addition to the algorithms. For example, job seekers and recruiters may potentially have biases that affect the way they use LinkedIn and interact with each other, which constitute human bias, as opposed to bias on the part of the algorithm. Given that measurements of algorithmic bias are often used as a pretext for algorithmic mitigations or to assign fault to the algorithm/platform, it makes sense to choose a metric that isolates the bias that is attributable to the algorithm. Later in the paper, we discuss the rationale behind separating bias on the part of the algorithm from other sources of bias and show that some commonly used algorithmic fairness metrics conflates them. The second criterion stipulates that any algorithmic bias metric should be robust to the fact that different groups may have different preferences or qualification distributions, which is a product of society and have been documented to exist. For example, men and women might make different education and career choice decisions, much of which predates their usage of LinkedIn but can potentially affect their behavior and outcomes on the platform. Furthermore, these preferences and qualifications are usually unobserved and difficult to quantify. We contend that a good algorithmic fairness metric should not impose restrictions on the preferences or qualifications that different groups may have. Later in the paper, we show that some common algorithmic fairness metrics miscategorize differences in qualification as algorithmic bias.

### 3.2 Equal opportunity for equally qualified

In this section, we define what "opportunity" and "qualification" entail in the context of a recommendation algorithm. Opportunities usually come in the form of a slot in a ranking or a treatment (e.g., a notification) that is allocated to some people but not others. For example, LinkedIn's PYMK algorithm gives members opportunities to connect with others by recommending them to other members in the form of a ranked list, the Feed algorithm gives members opportunities to disseminate potentially useful information by surfacing them in other members' Feeds, and the InstaJob algorithm enables members to be one of the first to apply for certain jobs by notifying them that a relevant job has just been posted[10].

Note that an *opportunity* is different from an *outcome*. An opportunity is an offering over which the algorithm has full control, e.g., recommending one member to another, sending a member a notification, etc. In contrast, an outcome is the action(s) taken in response to the opportunities that the algorithm confers and requires the voluntary participation of members (e.g., sending/accepting a connection invite, applying to a recommended job, etc.). For example, PYMK is a ranking algorithm that recommends potential connections to members. While LinkedIn offers an opportunity to a candidate by recommending him/her to a viewer, it is ultimately up to the viewer and the candidate to decide whether they want to connect. In this case, whether a connection resulted from the recommendation is the outcome of the recommended candidate. In general, a platform has full control over the opportunities it disseminates but cannot guarantee outcomes that require the participation of members.

We focus on two types of pointwise recommendation algorithms: classification and ranking. A canonical classification algorithm distributes a binary treatment, where all candidates whose scores fall above a common threshold receive the treatment, and those whose scores fall below the threshold do not receive the treatment. In this case, the opportunity is the treatment and it is being distributed by the candidates' scores. A ranking algorithm allocates opportunities in the form of slots in a ranked list with infinite scroll. Higher slots are considered better opportunities because they have a higher probability of being seen by the viewer, who may not scroll down far enough to see lower ranked candidates. Candidates are ranked in decreasing order of their predicted scores. For both pointwise classification and ranking algorithms, candidates with the same score will receive the same opportunity, on average[11]; therefore, equal opportunities translate into equal scores.

The criteria a pointwise recommendation algorithm uses to determine a candidate's qualification for an opportunity is stipulated by its objective or training labels, where candidates with higher predicted objective values, and hence higher scores, are considered more qualified/relevant. For example, InstaJob scores the "qualification" of candidates based on their predicted value of the following objective

$$Y_i = \alpha * 1(applied\ to\ job\ \&\ received\ no\ recruiter\ attention)_i \\ + 1(applied\ to\ job\ \&\ received\ recruiter\ attention)_i$$

For simplicity, we will refer to the predicted value of the algorithm's objective as the *predicted outcome*, $E[Y_i|X_i]$. Ex-post (after the algorithm has been deployed and recommendations have been made), we observe the actions taken with respect

---

[10] Note that the information being surfaced through LinkedIn's algorithms (PYMK, Feed, InstaJob, etc.) can be found even if it was not surfaced by the algorithm. For example, a member can discover other members through LinkedIn's search function, and they are not restricted to just those recommended by PYMK.

[11] In ranking algorithms, candidates with the same scores from the same query will receive the same opportunity (same rank), in expectation, as only one candidate can occupy a single slot. In classification algorithms, candidates with the score will receive the same treatment.

to the recommendations and each candidate's *realized outcome*. In the case of InstaJob, the realized outcome is 0 if the candidate did not apply to the job after being notified about it, $\alpha$ if the candidate applied to the job but did not receive recruiter attention, and 1 if the candidate applied to the job and received recruiter attention. Note that the realized outcome is the very outcome that the scores are designed to predict, and it is a realization of a random variable that is parameterized by the candidate's true qualification, $Y_i \sim F(.;q_i)$.

Using the notations introduced above, candidates $i$ and $j$ are equally qualified if $q_i = q_j$. Therefore, we can mathematically express the fact that an algorithm to give *equal opportunity for equally qualified candidates* as follows

$$q_i = q_j \rightarrow s(X_i) = s(X_j)$$

Conversely, a violation of this fairness notion can be expressed as follows, without loss of generality,

$$q_i = q_j \rightarrow s(X_i) > s(X_j)$$

In this case, candidate $i$ and $j$ are equally qualified, but the algorithm over-estimated the qualification of $i$ relative to that of $j$. And since higher scores translate into better opportunities, the algorithm is biased against $j$ relative to $i$.

Bias may creep into the algorithm in many ways. For example, if the training data is severely unbalanced, then the algorithm may be learning patterns predominantly from the majority group and in turn applying the same patterns to score all candidates during online model deployment; this can be problematic if minority groups do not follow the same mapping between features and outcome as the majority group. Another potential source of bias is non-representative training data; i.e., the data on which the model is trained is not representative of the population for which the model is making predictions. In this paper, we will focus on detection and quantification of bias and abstract from the source of the bias.

### 3.3 Detecting bias in the presence of unobservables

In this section, we will discuss how to empirically detect the algorithmic bias discussed in the previous section by drawing from the Economics literature and Gary S. Becker's seminal work, The Economics of Discrimination (1957). Becker's contribution, commonly known as the Outcome Test, remains an integral part of the Economics literature on detecting taste-based discrimination against a group in situations such as employment and policing, and more recently has been applied to detect algorithmic bias [2]. In the context of labor market discrimination, taste-based discrimination refers to discrimination against a group of candidates based on preferences that are unrelated to their qualifications[12]. For example, an employer is said to be engaging in taste-based discrimination against females if he/she has a preference against female candidates that is unrelated to their job performance.

A common challenge in determining whether an agent is engaging in taste-based discrimination is that not all factors used by the agent in his/her decision making are observed by the auditor. For example, the auditor may observe an employer hiring disproportionately more men than women relative to the applicant pool, but this may be because the job requires specific skills that are disproportionately more likely to be held by male applicants than by female applicants. If the auditor does not know all the skills and their relative importance for the job or is not able to assess candidates' skillset to the level of detail as the employer, then he/she cannot plausibly conclude that the employer is discriminating based on gender simply because the gender distribution of candidates who received an offer does not match that of the applicants. In other words, it is entirely feasible that the gender distribution of the most qualified candidates is different than that of the applicant pool.

---

[12] Taste-based discrimination stands in contrast to statistical discrimination where the agent does not have preferences over protected categories beyond their the signal they carry about the candidate's qualification.

The Economic literature circumvents the issue of unobservables with the following insights: a candidate's outcome is a realization of his/her true qualification; if two candidates are equally qualified, then they should realize equal outcomes, on average.

$$q_i = q_j \rightarrow E(Y_i|q_i) = E(Y_j|q_j)$$

Outcome here refers to the very outcome the candidate selection process is designed to predict. In the hiring example, a reasonable outcome the candidate selection process is designed to predict would be on-the-job performance. If two candidates are indeed equally qualified for a job, then they should perform equally well on the job if hired, in expectation. Conversely, if we see one candidate outperforming another, on average, it is evidence that the one with the higher average performance is more qualified.

$$E(Y_i|q_i) > E(Y_j|q_j) \rightarrow q_i > q_j$$

Looking at the outcomes instead of qualifications reduces the problem from a high dimensional feature space of qualifications to a one-dimensional space of outcomes. Similarly for algorithms, having an algorithmic bias metric that depends on candidate features can be potentially unwieldy, especially if historical feature data used to generate the scores are not available. In comparison, it is much easier to work with realized outcomes, which is usually stored for future model training.

### 3.4 Infra-marginality

Taste-based discrimination amounts to applying different standards to different groups, with the group being discriminated against subjected to a higher standard of qualifications. Simply comparing the average outcomes between groups is not informative of whether different groups are being subjected to different standards due to an issue that Economists call infra-marginality. We will illustrate the concept of infra-marginality with the following toy example. Consider an employment context where there are two groups of candidates, $G_i \in \{X, O\}$. The distribution of true qualification for each group is illustrated in in Figure 1. There is an equal mass of candidates from each group

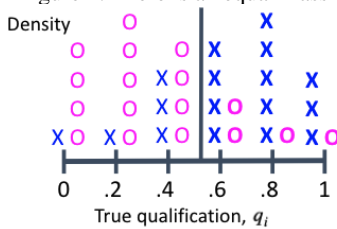

Figure 1: Distribution of true qualifications by group

Candidates' true qualifications are observable to the employer but not to the auditor. The employer hires according to the following rule: $h_i = 1(q_i \geq 0.6)$. Without loss of generality, assume candidates accept employment if offered. A candidate's outcome, $Y_i \sim Bernoulli(q_i)$, is a realization of a Bernoulli random variable parameterized by his/her true qualification, $q_i$, and is only observed if the candidate is hired. In this example, the employer is indeed giving equal opportunity to equally qualified candidates by applying a uniform hiring rule to both groups with respect to their true qualifications.

$$q_i = q_j \rightarrow T_i = T_j = 1(q \geq 0.6) \; \forall G_i, G_j$$

But if we calculate the average outcome of candidates who are hired for each group, we get

$$E(Y_i|h_i = 1, G_i = O) = \frac{0.6 * 2 + 0.8 + 1}{4} = 0.75$$

$$E(Y_i|h_i = 1, G_i = X) = \frac{0.6 * 4 + 0.8 * 5 + 1 * 3}{12} = 0.78$$

Through this example, we see that there can be differences between average group outcomes conditional on hiring even if the employer is being fair. This is because the average outcome is not only a function of the standard the employer applies to each group, but also of the distribution of true qualification that is unobserved to the auditor. To show this, we change the distribution of true qualifications to that in Figure 2 without changing the employer's decision rule, which is still fair, and this yields a different set of average outcomes.

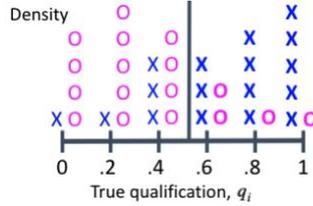

Figure 2: Distribution of true qualifications by group (different distribution from that of Figure 1 but same hiring standards)

$$E(Y_i|h_i = 1, G_i = O) = \frac{0.6 * 2 + 0.8 + 1}{4} = 0.75$$

$$E(Y_i|h_i = 1, G_i = X) = \frac{0.6 * 3 + 0.8 * 4 + 1 * 5}{12} = 0.83$$

The example above shows that group average statistics conditional on treatment varies with the underlying distribution of true qualification of each group, which is usually unobserved. Infra-marginality states that differences in average statistics is not informative of the fairness of the decision rule because it is confounded by differences in the distributions of true qualification, which is often unobserved.

### 3.5 Fairness condition

The Economics literature on discrimination circumvents the issue of infra-marginality by testing for equivalence of marginal outcomes between groups rather than that of average outcomes. The marginal outcome of a group is defined as the outcome of the least qualified candidate to receive a particular treatment, and because outcomes are realizations of one's true qualification, the marginal outcome of a group effectively traces out the implicit qualification standard being applied to the group for receiving that treatment.

In our toy example from the previous section, the marginal outcome of a group is the realized outcome of the least qualified candidate to get hired (receive treatment) from that group. In both examples from the previous section, the marginal outcomes are equal across groups, despite differences in the underlying qualification distribution, which is consistent with the fact that the decision rule is indeed giving equal opportunity to equally qualified candidates, regardless of their group membership.

$$E(Y_i|G_i \in O, q_i = 0.6) = 0.6$$

$$E(Y_i|G_i \in X, q_i = 0.6) = 0.6$$

Indeed, it is clear from Figures 1 and 2 that the marginal outcome is invariant to changes to the underlying qualification distribution.

To further illustrate how marginal outcomes are informative of bias, consider the following variation of the toy example from the previous section.

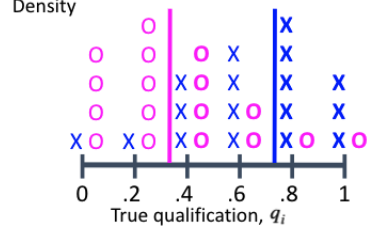

Figure 3: Distribution of true qualifications by group

In the example illustrated by Figure 3, the decision rule is biased as it distributes treatment to candidates from group $O$ with $q_i \geq 0.4$ but only distribute treatment to candidates from group $X$ if their true qualification is at least 0.8. Specifically, for candidates whose $q_i \in \{0.4, 0.6\}$, those who are equally qualified get different opportunities dependent on their group membership.

$$q_i = q_j \to T_i = 1 \neq T_j = 0 \text{ where } G_i = O, G_j = X$$

Calculating the marginal outcomes for each group reveals that the two groups indeed have different marginal outcomes, which is consistent with the fact that the decision rule is biased.

$$E(Y_i | G_i \in O, q_i = 0.4) = 0.4$$

$$E(Y_i | G_i \in X, q_i = 0.8) = 0.8$$

Note that the expected outcome of the least qualified candidates from each group traces out the effective qualification bar applied to each group for receiving treatment, which should be the same if the same standard is applied to all groups.

In general, the test for whether equal opportunity was given to equally qualified candidates, regardless of their group membership, amounts to testing the following null hypothesis

$$H_0: E(Y_i | i \in M^{g_1}) = E(Y_i | i \in M^{g_2}) \ \forall g_1, g_2 \in G$$

where $M^g$ denote the set of marginal candidates from group g.

### 3.6 Applying the Outcome Test to algorithms

In the context of an algorithm, the definition of a "marginal candidate" is specific to the opportunity being offered. For a classification algorithm, all candidates with scores above a common threshold, $\bar{S}$, receive the same opportunity, $T_i = 1(s_i \geq \bar{S})$. And because higher scores predict higher outcomes, on average, the marginal candidates consist of those whose scores are at the threshold, i.e., the candidates that the algorithm deems to be just qualified enough to receive treatment. In most cases, the number of candidates whose score is exactly the threshold score is too small to lend any statistical power to the test; therefore, we pick the set of marginal candidates to be those whose scores fall within a score bin that is right above the threshold[13].

---

[13] In the case where the score bin right above the threshold do not constitute the score segment with the lowest outcomes, on average, of the scores above the common threshold, it means that outcomes are not monotonically increasing in scores. This points to a problem in the

$$M = \{i : s_i \in [\bar{S}, \bar{S} + \epsilon]\}$$

Note that we usually do not observe the outcome of candidates who do not receive the treatment and therefore do not extend the bin to encompass those with scores below the threshold[14]. The the fairness condition can be directly applied to pointwise classification algorithms as follows,

$$H_0 : E(Y_i | i \in M, i \in g_1) = E(Y_i | i \in M, i \in g_2) \ \forall g_1, g_2 \in G$$

For a pointwise ranking algorithm, the opportunity being offered is a slot in a ranked list, where each slot constitutes a different opportunity. Given that candidates are ranked in decreasing order of their scores, a small change in a candidate's score can potentially result in a different opportunity (slot). In this case, each value the score takes constitutes a different margin, and we will test for equal (marginal) outcomes at each score value. The fairness condition for pointwise ranking algorithms, where the candidates being ranked belong in protected categories, is as follows.

$$H_0 : E(Y_i | s_i = s, i \in g_1) = E(Y_i | s_i = s, i \in g_2) \ \forall g_1, g_2 \in G, s \in \Omega$$

where $\Omega$ denotes the common support for scores between groups.

## 4 EMPIRICAL APPLICATION: DETECTING AND QUANTIFYING ALGORITHMIC BIAS

In this section, we apply our proposed test to two pointwise algorithms used at LinkedIn, a classification algorithm and a ranking algorithm, to test for algorithmic fairness with respect to the gender of the candidate being ranked[15]. We also conduct counterfactuals to quantify the magnitude of the bias in product metrics, that enables easier decision making on whether the bias is large enough to warrant mitigation. As of the writing of this paper, LinkedIn has operationalized bias detection and its mitigation in all its AI systems[16].

### 4.1 InstaJob

InstaJob is an algorithm that notifies LinkedIn members (candidates) regarding certain newly posted jobs that they may be interested in applying to[17]. We present and audit a old version of the InstaJob algorithm with respect to candidate gender, which takes the following categories: male, female, unknown-gendered. LinkedIn currently supports the measurement and mitigation of potential biases as described in the analysis. The algorithm uses features about the job posting and the candidate, $X_i$, to predict the likelihood that the candidate will respond to the notification by applying for the job and that the application will receive attention from the recruiter. Specifically, the algorithm scores candidates based on the predicted value of the following objective

$$Y_i = \alpha * 1(applied\ to\ job\ \&\ received\ no\ recruiter\ attention)_i$$
$$+ 1(applied\ to\ job\ \&\ received\ recruiter\ attention)_i$$

---

classification algorithm that goes beyond algorithmic fairness, because it means that giving the treatment to all those whose scores lie above a common threshold is not the outcome-maximizing allocation scheme.

[14] For example, if the treatment is a notification and it's only sent out to candidates with a score above a common threshold, then we do not observe how candidates with scores below the threshold would have behaved had they received the notification.

[15] This methodology can be applied to other groupings; however, we are unable to analyze or draw conclusions regarding other groupings due to the limited amount of data LinkedIn possesses regarding such groupings.

[16] https://engineering.linkedin.com/blog/2022/a-closer-look-at-how-linkedin-integrates-fairness-into-its-ai-pr

[17] Job seekers can still see and apply to the job on LinkedIn even if they do not receive an InstaJob notification regarding the job.

where $i$ is the candidate-job subscript, and $\alpha \in (0,1)$ is a (proprietary) scalar parameter that determines how the algorithm rewards applications with recruiter attention relative to those without. A notification is sent for all candidate-job pairs with a score above a common threshold, $\bar{S}$

$$T_i = 1(s_i \geq \bar{S}) \text{ where } s_i \equiv E(Y_i|X_i)$$

Figure 4 plots average realized outcomes by gender for each score deciles, where the realized outcomes are the very outcomes that the scores are designed to predict. Compared to female candidates with similar scores (same score decile), male candidates systematically realize higher outcomes, including in the marginal decile (the decile right above the threshold score, $\bar{S}$). The values of Y-axis are redacted for privacy reasons.

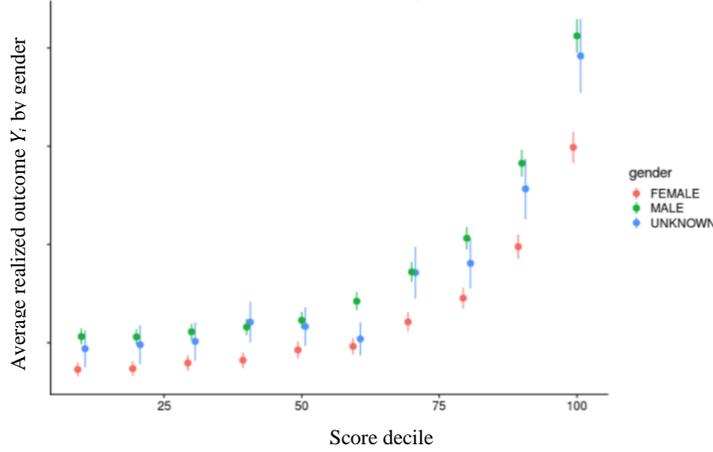

Figure 4: Average realized outcome by group for each score decile

The fairness condition for classification algorithms stipulates that the algorithm is giving equal opportunity to equally qualified candidates if we apply the same effective qualification standard to all groups, which is delineated by their marginal outcomes or the outcomes of the least qualified candidates to receive the treatment from each group. Therefore, we test the following fairness condition,

$$H_0: E(Y_i | i \in M, i \in g_1) = E(Y_i | i \in M, i \in g_2) \ \forall g_1, g_2 \in \{M, F, U\}$$

where $M$ is the set of marginal candidates. Specifically, we divide the scores of those who received treatment into deciles and test the fairness condition below for the marginal decile (the decile of scores right above the common threshold, $\bar{S}$)[18]. In our empirical implementation, we test the fairness condition by estimating the following linear specification on the marginal score decile, $i \in M$

$$Y_i = \alpha + \beta_1 * 1(MALE)_i + \beta_2 * 1(Unknown)_i + \gamma * s_i + \epsilon_i \quad (1)$$

where the score term controls for variations of scores within the marginal decile to help isolate the gender differences in marginal outcomes from variations in marginal outcomes due to the thickness of the score bin, $M$. The linear specification can be easily extended to a more flexible functional form to capture gender differences in marginal outcomes. Figure 5 plots $\widehat{\beta_1}$, $\widehat{\beta_2}$ from estimating specification (1) for each score decile.

---

[18] Deciding on the thickness of the marginal score bin comes down to a tradeoff between estimation precision (minimum detectable effect) and the variation in scores introduced by the width of the score bin.

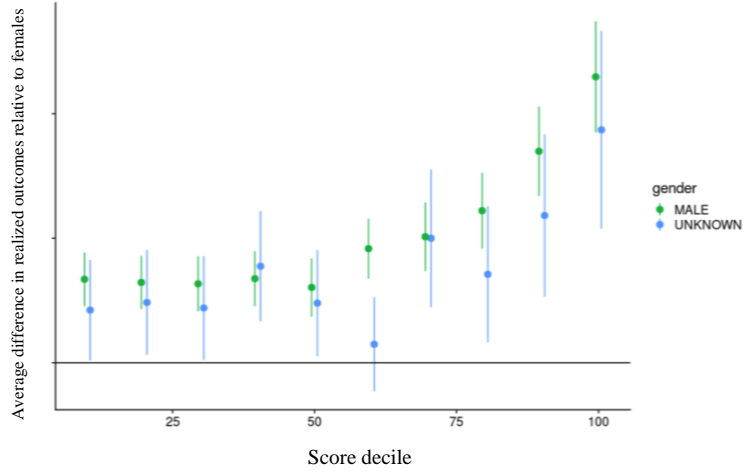

Figure 5: Difference in average realized outcome relative to females, controlling for score, for each score decile

A positive and statistically significant $\widehat{\beta_1}$ indicates that the algorithm systematically under-predicts the realized outcome of male candidates relative to that of female candidates, even after controlling for score variations within the score decile. And given that realized outcomes is a realization of the candidate's true qualification, it follows that the algorithm is under-predicting the true qualification of one group relative to another. Subjecting males and females to a common score threshold when the scores in the marginal decile underpredicts males' true qualifications relative to that of females implies that males and females are subjected to different qualification standards when it comes to treatment allocation. If the score distribution is sufficiently dense, this means that some male candidates who did not receive a notification were indeed more qualified than some female candidates who did receive a notification.

Next, we quantify the bias between candidates in terms of the number of notifications sent and in terms of the objective metric of the algorithm, $Y_i$. We do so with the following thought experiment: what if we eliminate the gender bias detected in (1) while holding the number of notifications sent for each job, $N_j$, at its previous level? More specifically, for a job, $j$, with $N_j$ notifications sent, we

1. Record the number of notifications sent to each group by the (biased) algorithm, $N^g = \sum_j N_j^g$, $g \in \{M, F, U\}$, where $N_j^g$ is the number of notifications for job $j$ that went to group $g$, and the total realized outcome metric, $Y = \sum_{i \in T} Y_i$, where $Y_i$ is the realized outcome of treated candidate-job pair $i$[19]. Note that, in this case, notifications were allocated using the following rule, $T_i = 1(s_i \geq \bar{S})$.
2. Estimate $\hat{Y}_i$ for each candidate-job pair using the estimates of (1). If the candidate-job pair has a score above the threshold, $s_i \geq \bar{S}$, then use the estimates of (1) from the corresponding score decile. If the candidate-job pair has a score below the threshold, $s_i < \bar{S}$, then estimate $\hat{Y}_i$ using the estimates of (1) from the marginal score decile, as the marginal score decile is the closest approximation, compared to higher score deciles, to the candidate-job pairs with scores below the threshold.

---

[19] If the notification for candidate-job pair $i$ resulted in the candidate applying to the job and receiving recruiter attention, then $Y_i = 1$; if it only resulted in the candidate applying to the job and his application did not receive recruiter attention, then $Y_i = \alpha$; and if it did not result in the candidate submitting an application, then $Y_i = 0$.

3. Record the number of notifications sent to each group under the bias-corrected counterfactual, $N^g_{bias-corrected} = \sum_j N^g_{j,bias-corrected}$, $g \in \{M, F, U\}$, where the notifications for job $j$ are allocated to the top $N_j$ candidates with the highest $\hat{Y}_i$'s for that job[20], and the total (predicted) outcome metric under this counterfactual, $\hat{Y} = \sum_{i \in N_{bias-corrected}} \hat{Y}_i$. In this counterfactual, we account for the gender bias in the original scores by using $\hat{Y}_i$'s, instead of $s_i$'s, to allocate notifications; in other words, we are distributing notifications for each job to candidates with the highest predicted outcome, after accounting for gender-miscalibration in the original scores.

4. Calculate the percent change in number of notifications sent to each group and total outcome metric between the two scenarios (biased vs bias-corrected counterfactual)

$$N^g_{\%\Delta} = \frac{N^g_{bias-corrected} - N^g}{N^g} * 100$$

$$Y_{\%\Delta} = \frac{\hat{Y} - Y}{Y} * 100$$

Under the bias-corrected counterfactual, the number of notifications sent to male and unknown-gendered candidates increased and the number of notifications sent to female candidates decreased, compared to their respective numbers under the original (biased) formulation. The exact percent change in the number of notifications sent to each gender between the original formulation and the bias-corrected counterfactual is redacted for privacy reasons. The total number of notifications sent, per job, stays constant by design. The total realized outcome increased by almost 1% under the bias-correct counterfactual compared to the original biased formulation, despite holding constant the number of notifications sent. This is because we are correcting for gender-bias in the notification allocation scheme by trading less qualified candidates for more qualified candidates in the counterfactual, which in turn yields a higher outcome metric. Note that the change in outcomes is small compared to the change in the number of notifications sent for each group because the candidates whose treatment assignment changed under the bias-corrected counterfactual are from the marginal decile and are the least likely to realize a positive outcome (apply for a job and/or receive recruiter attention) out of those who were treated. Because the ultimate goal of the algorithm is to find the best candidate-job matches, which is captured in higher outcomes, $Y_i$, quantifying algorithmic bias in terms of percent change in the outcome presents an easier-to-interpret bias metric than quantifying the bias in terms of notifications sent to each group, which may or may not result in improvements outcome of interest.

## 4.2 PYMK

In this section, we present the application of our proposed test for algorithmic bias to audit an old version of the PYMK algorithm. LinkedIn currently supports the measurement and mitigation of potential biases as described in the analysis. PYMK is a pointwise ranking algorithm with an objective metric, $Y_i$, that is a function of whether a connection invite was sent, accepted, and some downstream interaction metrics between the connected members. The actual formula is proprietary and therefore redacted. Candidates are scored based on their predicted value of the objective metric, $s_i = E(Y_i|X_i)$, and ranked in decreasing order of the scores. Figure 6 plots average realized outcomes by gender for each score deciles, where the realized outcomes are the very outcomes that the scores are designed to predict. Compared to male candidates with similar scores (same score decile), female candidates systematically realize higher outcomes. The values of Y-axis are redacted for privacy reasons.

---

[20] Note, that we are redistributing notifications for a job to the same set of candidates that were scored for that job under the original algorithm. The counterfactual would not allocation a job notification to a candidate if the candidate-job pair was not scored under the original algorithm.

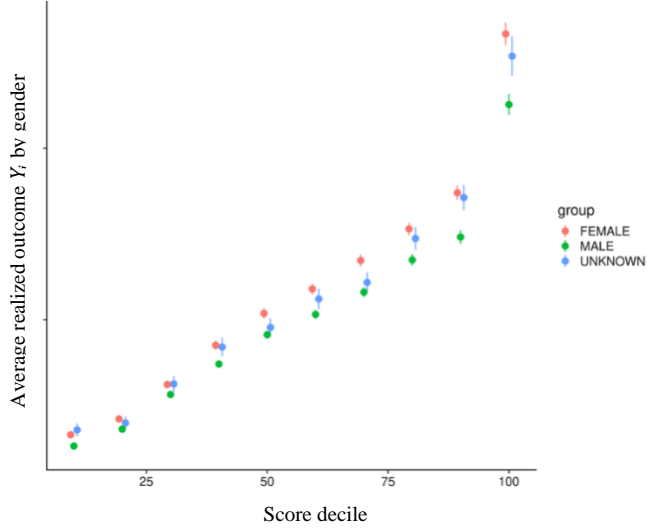

Figure 6: Average realized outcome by group for each score decile

Unlike the fairness condition for pointwise classification algorithms, the fairness condition of a pointwise ranking algorithm is tested on the entire common support of scores, Ω, rather than on just the marginal score. The fairness condition stipulates that those assigned the same score by the algorithm, $s \in \Omega$, should go on to realize the same outcomes, on average, regardless of protected categories. Violation of this fairness condition anywhere along the common support of scores points to algorithmic bias. Because conditional on single score value does not lend enough statistical power to the test, we first divide the common support of scores into deciles[21]. For each score decile, $S_d$, we estimate the following linear specification of the fairness condition,

$$Y_i = \alpha + \beta_1 * 1(Female)_i + \beta_2 * 1(Unknown)_i + \gamma * s_i + \epsilon_i \qquad (2)$$

where the score term controls for variations of scores within the score decile to help isolate the gender differences in outcomes from variations in outcomes due to the thickness of the score decile, $S_d$. The estimate of coefficient $\beta_1$ can be interpreted as the average difference in realized outcomes between male and female candidates conditioning on their scores within a particular decile. In other words, the empirical specification approximates comparing candidates who share a single score value by separately testing the fairness condition on each score decile and by linearly controlling for score variations within each decile. The linear specification can be easily extended to a more flexible functional form to capture gender differences in conditional outcomes.

Figure 7 plots the $\widehat{\beta_1}$, $\widehat{\beta_2}$ from estimating specification (2) for each score decile. Note that $Y_i$ is the realized outcome of PYMK's objective, which is different than that of InstaJob.

---

[21] The number of bins to split the score distribution into amounts to a tradeoff between the power of the test, which increases with the number of observations in a bin, and the precision of the conditional expectation, where conditioning on a score bin is the empirical approximation of conditioning on a single score value in the fairness condition.

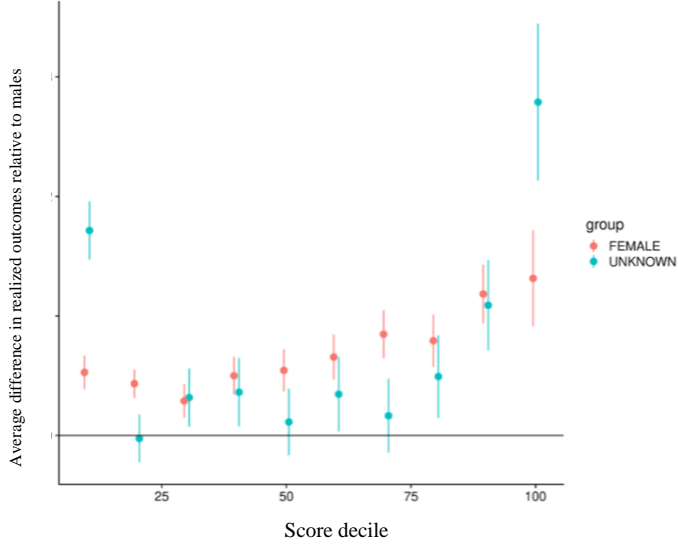

Figure 7: Difference in average realized outcome relative to females, controlling for score, for each score decile

A positive and statistically significant estimate for $\widehat{\beta_1}$ indicates that male and female candidates who were deemed to be equally qualified by the algorithm, and hence assigned the same score, actually go on to realize different outcomes—the very outcome that the scores are designed to predict. This indicates that the algorithm is under-estimating the qualification of female candidates relative to that of male candidates, which can potentially result in mis-ranking between genders.

We quantify the magnitude of the algorithmic bias in terms of the change in the average rank of each group through the following thought experiment: how would the average rank of each group change if we correct for the gender bias detected in (2) along the entire score distribution? Specifically, we

1. Calculate the average rank of impressed candidates from each group, $\overline{r^g} = \frac{1}{|g|} \sum_{i \in g} r_i$, generated by the original algorithm, which ranks candidates in decreasing order of their scores.
2. Estimate $\widehat{Y_i}$ for each candidate using the estimates of (2) for the appropriate score decile.
3. Simulate the counterfactual where candidates, within each query, are instead ranked in decreasing order of $\widehat{Y_i}$ and record the bias-adjusted average rank for each group, $\overline{r^g_{bias-adjusted}} = \frac{1}{|g|} \sum_{i \in g} r_{i,bias-adjusted}$. Only the top $N_j$ candidates for each query are used to calculate the bias-adjusted average rank, where $N_j$ is the number of impressed candidates for query $j$.
4. For each group, calculate the change in average rank between the bias-adjusted counterfactual and the original ranking.

$$\overline{r^g_\Delta} = \overline{r^g_{bias-adjusted}} - \overline{r^g}$$

By re-ranking impressed candidates based on $\widehat{Y_i}$, we effectively accounted for the gender-based algorithmic bias in the original scores. In the counterfactual simulation where we remove between-gender miscalibration, male candidates moved up (higher in the ranking), whereas female and unknown-gendered candidates moved down, on average, compared to their respective positions in the original rankings. The exact change of each gender's average ranking between the original rankings and the bias-corrected counterfactual rankings is redacted for privacy reasons. These results are consistent with our finding that the algorithm is under-scoring female and unknown-gendered candidates relative to their male counterparts, where candidates returned in a single query are ranked in decreasing order of their scores.

## 5 DISCUSSION

### 5.1 Algorithmic bias vs human bias

We introduce a clean way to distinguish between algorithmic bias and human bias for recommendation algorithms, where the outcome is determined by the interplay between the algorithm and human agency, and we illustrate the delineation between the two sources of bias by using LinkedIn's InstaJob as an example. As an online platform, LinkedIn hosts a variety of algorithms that make recommendations to members and customers. However, one's experience on LinkedIn is not only determined by LinkedIn's algorithms but also the way in which members interact with other members, from which the algorithms learns and makes personalized recommendations. When it comes to ensuring fairness on LinkedIn, it is important to recognize that there are elements beyond LinkedIn's control and inequities that cannot be fixed by the algorithm. When taking steps to detect, measure, and mitigate bias, it is helpful to distinguish between what the algorithm can and cannot fully control. We restrict our definition of algorithmic bias to the bias for which the audited algorithm is solely responsible, on top of any potential human bias or pre-existing differences between groups. Likewise, we restrict our definition of human bias as the bias that human actors exercise when evaluating candidates that are available to them.

InstaJob is an algorithm that recommends certain newly posted jobs to potential job seekers through a notification. When a job is posted, InstaJob scores potential candidate-job pairs based on its predicted value of an outcome metric that is a linear combination whether the candidate will apply for the job if notified and whether the application will receive recruiter attention conditional on applying. If the candidate-job pair's score lies above a certain common threshold then the algorithm sends a notification to the candidate alerting him/her that of the job posting.

One important feature about the InstaJob algorithm is that, like most algorithms on LinkedIn, it only makes recommendations, and the job seeker is the final decision maker in whether to apply for the job. In the case of InstaJobs, human bias can manifest in two ways.

- Job seeker bias - the job seeker can exercise self-censoring when deciding whether to apply to a job. There is evidence in the women self-censor more than similarly qualified men when it comes to applying for jobs [9].
- Recruiter bias - after job seekers have applied to a job, recruiters can be biased in deciding whose application they want to move forward.

Note, that these are biases that LinkedIn users can manifest upon getting a recommendation from LinkedIn (or in response to an action taken with respect to a recommendation from LinkedIn).

On the other hand, recommendation algorithms operate as a marketplace matchmaker that facilitates the search and discovery of relevant information. For example, InstaJob deliver values to job seekers and recruiters by matching those with mutual interest. InstaJob is not able to dictate actions on the part of job seekers or recruiters. Instead, it is designed to learn the preferences of both parties, predict their actions, and to make recommendations as to maximize the outcome of interest, which is specified by its objective. Therefore, we define the bias of a recommendation algorithm relative to its given objective. We say that a recommendation algorithm is biased against a group if it under-predicts the outcome of the group relative to that of other groups and this under-prediction leads to the mis-allocation of opportunities; in other words, the recommendation algorithm is biased if it does not give equal opportunities to equally qualified candidates, regardless of their group membership. In the case of InstaJobs, algorithm bias arises when the algorithm under-predicts one group's outcome relative to another's, which can potentially lead to the treatment being denied to better qualified candidates of the group against which the algorithm is biased in favor of less qualified candidates from another group.

We distinguish between algorithmic bias and human bias because they require different mitigations, and being able to identify the source of bias allows us to enact the correct mitigation. In the case of a biased InstaJobs algorithm, algorithmic bias can easily be mitigated by fixing the miscalibration in the scores, which will lead to a different set of recommendations. However, doing so will not be effective in mitigating human bias; in particular, giving or taking away a job notification would not change job seekers' inclination to engage in self-censoring or make biased recruiters less discriminatory. What could potentially be effective for mitigating human biases are product changes; for example, to mitigate self-censoring, which originates from a difference in subjective self-evaluation and how one would be evaluated by others, the platform might offer an objective evaluation of where a job seeker ranks in the pool of applicants through messages such as "you are in the top 10% of applicants"; to mitigate recruiter bias, we could highlight applicants' experience and education to draw attention away from their profile pictures or names, which contain information about their protected categories. Of course, these types of behavioral interventions would not be effective in mitigating algorithmic bias. The 2x2 table below illustrates the different mitigations necessary for different sources of biases.

| Example: Instajobs | Mitigation method | |
|---|---|---|
| | **Algorithmic mitigation**: fix the bias in scoring → different recommendations | **Behavioral/product mitigation**: e.g., surface more relevant information to minimize reliance on protected categories |
| **Algorithmic bias**: systematically underestimates one group's outcomes relative to that another's | ✓ | **Not effective** - behavioral or product interventions will not fix the bias in the algorithm's scores |
| **Human bias**: group-based discrimination not justified by qualifications<br>Job seekers: self-censoring<br>Recruiters: taste-based discrimination | **Not effective** - reallocating notifications will not mitigate self-censoring on the part of the job seekers or bias on the part of recruiters | **Examples of behavioral/product interventions**<br>● Job seekers: "you're in the top 10% of applicants"<br>● Recruiters: highlight more relevant information like education and experience |

(Source of bias — row axis)

Formally, we define and differentiate between algorithmic bias and human bias as follows. Algorithmic bias arises when the algorithm underestimating one group's qualifications (as defined by the algorithm's objective/labels) relative to another's, which leads to a violation of the fairness notion of equal opportunities for equally qualified. Human bias, on the other hand, is where human actors exhibit bias when making decisions over the choices with which they were presented. Because human actors are the final gatekeepers of actions, changes to the algorithm alone cannot mitigate all bias on the platform. The framework we introduce attributes bias to its source and illustrates the limitation of algorithmic mitigations in eliminating bias from the platform.

## 5.2 Algorithm's objective

So far, we have audited recommendation algorithms with respect to the objective they are given; this is because group differences in any outcome other than the algorithm's objective function can simply be due to the fact that the algorithm was not designed to make such predictions and therefore makes no guarantee with respect to them. However, we could think about a extending the framework for ensuring "fairness" in a way that encompasses fairness in the design of the algorithm's objective. An analysis of this expanded framework, however, are beyond the scope of this paper.

## 5.3 Limitations and context

In this section, we discuss the limitations of our analysis. First, the fairness metric we derived is a based on a very specific notion of fairness, equal opportunity for equally qualified candidates, which may run counter to other notion fairness or equity. Our main goal is to derive a fairness metric that embodies a particular fairness notion, rather than to arbitrate between different notions of fairness or equity. Second, our fairness metric, like most in the algorithmic fairness literature, measures the fairness of an algorithm at a point in time. Dynamics, such as those that arise from retraining the algorithm using outputs of the previous version of the algorithm, and its intersection with algorithmic fairness, is beyond the scope of this paper. Third, this paper focuses the on the measurement of algorithmic bias, rather than identifying its source, such as the training data or model features. But regardless of source of the algorithmic bias, it must manifest in the algorithm's score in order to affect the distribution of opportunities, which is dictated by scores. As such, we consider the algorithm's score a sufficient statistic of its Finally, we audit the algorithm with respect to the objective that they are given as the objective dictates how an algorithm assesses candidates' qualifications. Although beyond the scope of the current paper, an interesting future research direction would be to conduct a deeper examination into the fairness implications of the algorithm's objective.

## 5.4 Other common fairness metrics

In this section, we discuss a couple of common algorithmic fairness and their incongruence with the fairness notion of equal opportunity for equally qualified candidates. We do so by giving counterexamples of cases where the fairness metric in question mis-measures a fair decision rule as unfair and cases where it mis-measures an unfair scenario as fair.

### *5.4.1 Equalized odds*

Equalized odds stipulates that the score distribution must be independent of the protected category, conditional on the realized outcome.

$$s \perp G \mid Y$$

We constructed the following examples in a pointwise ranking setting (e.g., PYMK), where candidates are ranked by their scores in decreasing order, and the realized outcome is observed only if the candidate received an impression.

#### 5.4.1.1  False positive

Consider the following example, where the distribution of true qualifications for each protected category, $\{X, O\}$, is as follows.

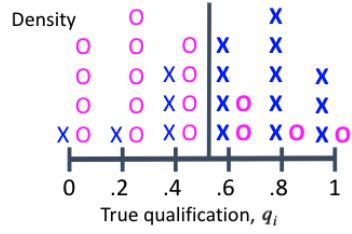

Each X or O represents a unit mass, and there are equal number of candidate from each group. For each query, a random subset of candidates are drawn from the above distributions of true qualifications and scored. Algorithm scores accurately predicts each candidate's true qualification, $s_i = q_i$, and candidates are ranked in decreasing order by their scores; if scores are tied, then a coin is flipped to determine the order. The viewer sends an invite to the candidate upon impression if the candidate has a true qualification greater than 0.6; i.e., $Y_i = 1(q_i \geq 0.6)$. Assume all candidates returned by the query receives an impression. Note that each candidate's true qualification is observable to the viewer but unobserved by the auditor.

In this example, the algorithm is indeed fair because it is giving equal opportunity to equally qualified candidates.
$$q_i = q_j \Rightarrow s_i = s_j, \quad \forall i \in X \,\&\, j \in O$$
However, the equalized odds fairness metrics mis-labels the algorithm as unfair; conditional on $Y_i = 1$, the score distribution is clearly not independent of $G_i$; same if we condition on $Y_i = 0$.

#### 5.4.1.2  False negative

Consider the following example, where the distribution of true qualifications for each protected category, $\{X, O\}$, is as follows.

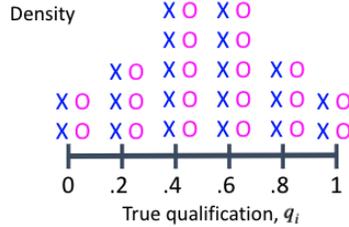

For each query, a random subset of candidates are drawn from the above distributions of true qualifications and scored, where the algorithm scores candidates using the following mapping; $s_i = q_i \; if \; i \in O$ and $s_i = (1 - q_i) \; if \; i \in X$. Candidates are ranked in decreasing order by their scores; if the scores are tied, then a coin is flipped to determine the order. The viewer sends an invite to a candidate upon impression with probability equal to the candidate's true qualification, $pr(Y_i = 1 \,|T_i = 1) = q_i$. Assume all candidates returned by the query receive an impression. Note candidates' inherent qualification is not observable to the researcher but it is observable to the viewer.

In this example, the algorithm is not fair because it's not giving equal opportunity to equally qualified candidates regardless of their protected attribute. In particular,
$$q_i = q_j \Rightarrow s_i > s_j \; for \; q \geq 0.6 \; and \; i \in O \,\&\, j \in X$$

$$q_i = q_j \implies s_i < s_j \text{ for } q < 0.6 \text{ and } i \in O \ \& \ j \in X$$

However, the equalized odds metrics mis-labels this algorithm is fair, because

$$pr(s_i = s \,|\, i \in O, Y_i = r) = pr(s_i = s | i \in X, Y_i = r) \ \forall s, r$$

### 5.4.1.3 Metric flaw

The problem with the equalized odds metric is that it does not distinguish between differences in groups' true qualification distributions and unfairness on the part of the algorithm. To see this, we can go back to the false positive counterexample and change the distribution of inherent qualifications, without changing how the algorithm scores candidates or the decision rule of the viewer; doing so will yield a different equalized odds metric. In the false negative counter-example, we saw that the equalized odds metric mistook different groups having the same distribution of inherent qualifications with fairness on the part of the algorithm. Therefore, violation of equalized odds is not informative of whether the algorithm is actually giving equal opportunity for equally qualified candidates.

### 5.4.2 Precision

Precision is a metric often used in the context of a pointwise binary classifiers that scores candidates based on their propensity to realized a binary outcome, $Y_i \in \{0,1\}$, and allocates the treatment to those whose predicted score is above a common threshold; i.e., $T_i = 1(s_i \geq \bar{S})$. A group's precision is defined as the proportion of the group that realized a positive outcome conditional receiving treatment.

$$Precision(G_i = g) = \frac{\sum_{i \in g} Y_i}{\sum_{i \in g} T_i}$$

When Precision is used as an algorithmic fairness metric, the fairness condition usually requires equal precision between groups; i.e.,

$$Precision(G_i = g_1) = Precision(G_i = g_2) \ \forall g_1, g_2 \in G$$

### 5.4.2.1 False positive

Consider the following example, where the distribution of true qualifications for each protected category, {X,O}, is as follows.

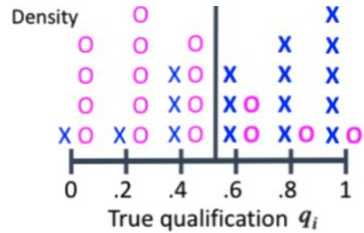

Each X or O represents a unit mass, and there are equal number of candidates from both groups. The algorithm scores perfectly predict each candidate's true qualification, $s_i = q_i$, and the binary treatment is allocated to those with scores above 0.6, $T_i = 1(s_i \geq 0.6)$. In this example, the algorithm is indeed fair because it is giving equal opportunity (treatment) to equally qualified candidates.

$$q_i = q_j \implies s_i = s_j \implies T_i = T_j, \qquad \forall i \in X \ \& \ j \in O$$

However, when we calculate group precision, we see that they have different precisions, which constitutes a false positive.

$$Precision(X) = \frac{0.6 * 4 + 0.8 * 5 + 1 * 3}{4 + 5 + 3} = 0.78$$

$$Precision(O) = \frac{0.6 * 2 + 0.8 * 1 + 1 * 1}{2 + 1 + 1} = 0.75$$

#### 5.4.2.2  False negative

Consider the following example, where the distribution of true qualifications for each protected category, {X,O}, is as follows.

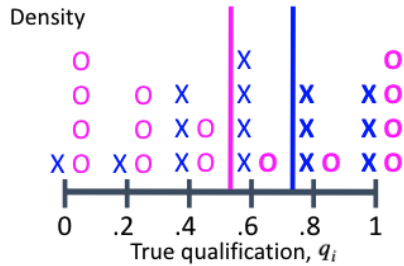

Each X or O represents a unit mass, and there are equal number of candidates from both groups. The algorithm scores map to each candidate's true qualification as follows, $s_i = q_i - 0.2 \ for \ i \in X$, $s_i = q_i \ for \ i \in O$. A binary treatment is allocated to those with a score above 0.6, $T_i = 1(s_i \geq 0.6)$.

In this example, the algorithm is not fair because it is underestimating the true qualification of candidates from group X while correctly estimating those of O. Although everyone is subjected to the same nominal threshold of $\bar{S} = 0.6$, this underestimation of one group but not another leads to different effective standards being applied to different group and to equally qualified candidates receiving different treatments:

$$q_i = 0.6 \ and \ i \in X \implies s_i = 0.4 \implies T_i = 0$$

$$q_i = 0.6 \ and \ i \in O \implies s_i = 0.6 \implies T_i = 1$$

However, when we calculate group precision, we see that they have the same precision, which mislabels this algorithm as fair.

$$Precision(X) = \frac{0.8 * 3 + 1 * 3}{3 + 3} = 0.9$$

$$Precision(O) = \frac{0.6 * 1 + 0.8 * 1 + 1 * 4}{1 + 1 + 4} = 0.9$$

#### 5.4.2.3  Metric flaw

The problem with Precision is that it conflates differences in the distribution of true qualifications between groups with algorithmic bias. As shown in the false positive example above, differences in group precision can exist even if the algorithm perfectly predict each candidates' inherent qualifications. The false negative example above shows that the differences in the distribution of inherent qualifications between groups can numerically offset actual algorithmic bias to produce equal group precisions even when there is clear algorithmic bias based on group membership.

## 6 CONCLUSION

In this paper, we derived at an algorithmic fairness metric for a two-sided platform from the fairness notion of giving equal opportunity for equally qualified candidates. We borrow principles from the Economic literature on discrimination that are used to detect taste-based discrimination in settings such as employment and policing and apply them to test for algorithmic bias in pointwise scoring classification and ranking algorithms. We demonstrate the application of our algorithmic fairness metric by auditing a classification algorithm and a pointwise ranking algorithm from LinkedIn with respect to gender. Furthermore, we discuss how to quantify the size of the bias in easy-to-understand product metrics that measure the actual impact on the end users. Finally, we introduce a framework for how to decouple algorithmic bias from human bias in a two-sided marketplace and discuss the incongruence of a couple of commonly used fairness metrics with our chosen fairness notion.